\RequirePackage{fixltx2e}
\documentclass[aip,jcp,reprint,showpacs,singlecolumn]{revtex4-1}

\usepackage{epsfig,amsopn}
\usepackage{graphicx}
\usepackage{amsmath,amssymb}
\usepackage{comment}

\usepackage{color}
\usepackage{placeins}
\newcommand{\be}{\begin{equation}}
\newcommand{\ee}{\end{equation}}
\newcommand{\bea}{\begin{eqnarray}}
\newcommand{\eea}{\end{eqnarray}}
\usepackage{dcolumn}
\usepackage{bm}
\usepackage{hyperref}
\usepackage{epsf}
\usepackage{subfigure}
\usepackage{epstopdf}%
\usepackage{amsfonts}

\begin{document}

\author{Michael Kilgour}
\affiliation{Chemical Physics Theory Group, Department of Chemistry,
University of Toronto, 80 Saint George St., Toronto, Ontario, Canada M5S 3H6}
\author{Dvira Segal}
\affiliation{Chemical Physics Theory Group,
Department of Chemistry, University of Toronto, 80 Saint George St., Toronto, Ontario, Canada M5S 3H6}

\title{Tunneling diodes under environmental effects}

\begin{abstract}
We examine the robustness of single-molecule tunneling diodes to thermal-environmental effects.
The diode comprises three fragments: two different conjugated chemical groups at the boundaries,
and a saturated moiety in between, breaking conjugation.
In this setup, molecular electronic levels localized on
the conjugated groups independently shift with applied bias.
While in the forward polarity a resonance condition is met, enhancing
conductance, in the reversed direction molecular electronic states shift away from each other,
resulting in small tunneling currents.
In the absence of interactions with a thermal environment
(consisting e.g. internal vibrations, solvent),
rectification ratios reach three orders of magnitude.
We introduce decoherence and inelastic-dissipative effects phenomenologically, by using the ``voltage probe" approach.
We find that when $\gamma_d \lesssim v$, with $\gamma_d$ the interaction energy of electrons with the environment
and $v$ the tunneling energy across the saturated link,
the diode is still highly effective,
though rectification ratios are cut down by a factor of 2-4 compared to the coherent limit.
To further enhance rectification ratios in molecular diodes
we suggest a refined design
involving four orbitals, with a pair of closely spaced states at each conjugated moiety.
\end{abstract}
\maketitle


\section{Introduction}

One of the simplest building blocks in ordinary semiconductor circuitry,
essential for realizing molecular-based electronics, is the rectifier (diode).
In fact, the Aviram-Ratner proposal for an organic molecular rectifier
largely initiated the field of molecular electronics in 1974 \cite{AR}. Since then,
molecular rectifiers have been fabricated
from thin films, self-assembled monolayers (SAMs)
\cite{Martin,Ng,Mujica,Metzger1, Metzger2},
and single molecules \cite{Mayor,Tao,latha13}.
However, in general reported rectification ratios were rather low.
Only recently have SAMs
\cite{Nijhuis09,Nijhuis11,Nijhuis13,Nijhuis15} and single molecule junctions \cite{Cuevas,latha15}
provided robust molecular rectifiers with
high conductance, and rectification ratios of two and even three orders of magnitude.

What is the mechanism of electron current rectification in molecular junctions?
The literature includes several constructions based on
molecules with an asymmetric backbone, different molecule-metal linkers at the two ends,
or electrodes of different materials \cite{MetzgerRev}.
Fundamentally, from the principles of quantum transport, it is apparent that
coherent conduction of noninteracting electrons,
as described in scattering theory by the Landauer formula, cannot
materialize rectification
when energy levels are fixed, independent of applied bias \cite{datta}.
Two necessary conditions should be simultaneously met for achieving the diode effect:
many body interactions should play a role, and a spatial asymmetry should be built into the junction,
see e.g. Refs. \cite{Braunecker,SB}, exemplifying charge and heat rectification.

In many proposals, however, the role of many body effects is not spelled out, and it is only
included at the level of mean-field, by assuming a screening interaction which creates
a voltage drop across the molecular bridge \cite{MetzgerRev,Nitzan}.
In particular, theoretical and computational investigations of tunneling diodes employ
molecules that are asymmetrically coupled to the electrodes.
Using model calculations or first-principle simulations,
it is assumed or demonstrated that the junction develops an asymmetric
potential profile under bias in the forward and backward direction, see e.g. Refs. \cite{Korn,Taylor,Cuevas,Batista}.
This in turn results in an asymmetric IV characteristics.
The development of a voltage drop {\it inside} the molecule, however, can be only justified
when many body effects (essentially electron-electron interactions) play a role in the junction \cite{commentMB}.

\begin{figure*}[htbp]
\vspace{0mm}
{\hspace{-12mm}\hbox{\epsfxsize=190mm \epsffile{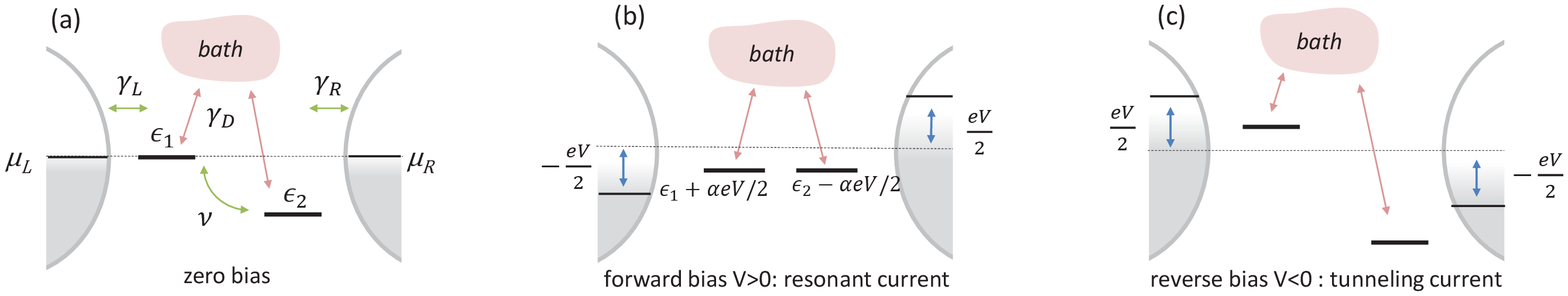}}} 
\caption{Illustration of the examined tunneling diode.
Displayed are levels (a) at zero bias, (b) forward bias,
providing a resonant situation beneficial for conductance,
and (c) under reverse bias when levels are shifted away from each other resulting in small tunneling currents.
}
\label{Fig1}
\end{figure*}

In this paper, we are concerned with single molecule rectifiers made of three fragments,
e.g., donor-$\sigma$-acceptor junctions.
While in their original proposal Aviram and Ratner had assumed vibrationally-assisted
intramolecular electron transfer  \cite{AR},
a related diode can be realized based on quantum tunneling,
as presented in Figure \ref{Fig1} and discussed in the next section.
Tunneling diodes were investigated for example
in Ref. \cite{vanderzant15}, where
a DFT + NEGF approach was employed for the study of coherent transport in
a three-fragment 1,2-bis(4-(phenylethynyl)phenyl), showing
a rectification ratio of $R\sim 1000$.
A similar approach (with a different backbone) was adopted in Ref. \cite{Ratnerpn15}, manifesting
$R\sim 100$. These studies and similar proposals
employ the Landauer formula for coherent conduction, with the additional assumption
of level shift under bias.

The objective of the present work is to examine the robustness of tunneling diodes
against environmental effects, when electrons on the junction suffer from decoherence,
inelastic effects, and energy dissipation due to their interaction with an ``environment",
consisting of degrees of freedom such as molecular vibrations, the solvent,
and other electrons. Naturally, since the rectification mechanism in tunneling diodes,
depicted in Fig. \ref{Fig1},
relies on the sharp contrast between resonant transmission
and deep tunneling, rectification ratios should deteriorate when
incoherent processes play a role.
However, an estimate of this effect for realistic molecular diodes has not yet been provided.

\section{Model and Method}

\begin{figure}[htbp]
\vspace{0mm}
\hspace{-0mm}{\hbox{\epsfxsize=80mm \epsffile{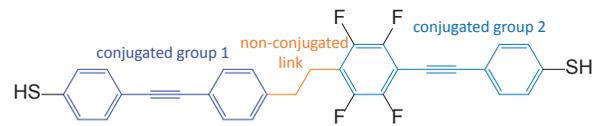}}} 
\hspace{-0mm}{\hbox{\epsfxsize=90mm \epsffile{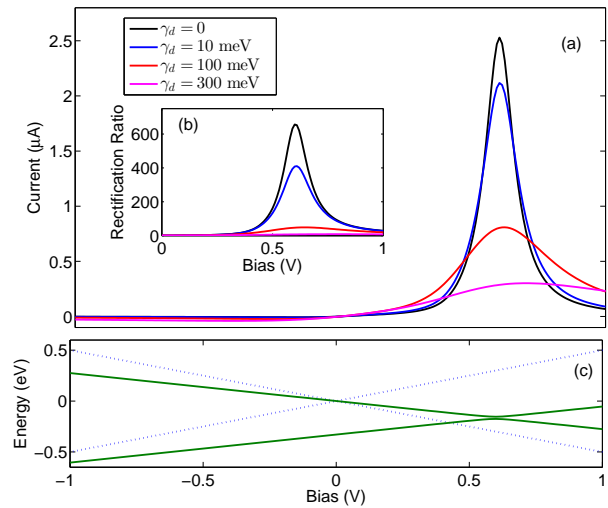}}} 
\vspace{-4mm}
\caption{
Tunneling diodes under inelastic effects with parameters corresponding to the
 1,2-bis(4-(phenylethynyl)phenyl)ethane molecule (top panel), serving as the backbone of the diode.
(a) IV characteristics of the junction,
(b) rectification ratio $R$, and (c)
level diagram showing the chemical potentials $\mu_{L,R}$ at the two leads (dashed lines)
 and the eigenenergies of $\hat H$, Eq. (\ref{eq:H})
(full lines).
Parameters here and throughout the paper
correspond to a two-state model of the molecule (top panel)
with $\epsilon_1=0$, $\epsilon_2=-0.329$, $v=0.0109$, $\gamma_{L,R}=0.03$, all in units of eV,
and $\alpha=0.55$ \cite{vanderzant15}, $T=298$ K.
$\gamma_d=0, 0.01,0.1, 0.3$ eV. 
}
\label{Fig2}
\end{figure}

We focus on recent proposals for highly rectifying tunneling diodes built from
 three-fragment molecular backbones
such as 1,2-bis(4-(phenylethynyl)phenyl)ethane junctions \cite{vanderzant15}, see Fig. \ref{Fig2}.
This asymmetric molecule includes conjugated groups at each end of the junction
and a weak tunneling link in the form of a saturated carbon bridge,
breaking conjugation.
IV characteristics of this system, as calculated with a DFT + NEGF approach, were
excellently matched with the two-site (HOMO and HOMO-1) model Hamiltonian
\bea \hat H=\left[  \begin{array}{cc}
 \epsilon_1  + \frac{1}{2}\alpha eV  & v\\
v & \epsilon_2-\frac{1}{2}\alpha e V \\
 \end{array}\right],
\nonumber
\label{eq:H}
\eea
%
with $\epsilon_{1,2}$ as the
conjugated groups' site energies at zero bias, $v$ a small tunneling energy between the units,
and $\alpha$ a phenomenological parameter which describes the fraction of the
voltage drop within the molecule, the response to the screening interaction.
At zero bias, $\epsilon_1=0$  and $\epsilon_2<0$.
Under forward bias defined as $V>0$,
a resonance situation can be met once $\epsilon_1+\frac{1}{2}\alpha eV \approx\epsilon_2-\frac{1}{2}\alpha eV$,
with the shifted levels buried within the bias window, resulting in high currents.
Under the reversed operation only tunneling currents contribute in
coherent scenarios, see Fig. \ref{Fig1}.
Coherent transport can be described by the  Landauer formalism
\cite{datta}.
We construct the metal-molecule hybridization matrices with coupling strengths $\gamma_{L,R}$,
\bea &&\hat \Gamma_{L}=\gamma_L\left[ \begin{array}{cc}
              1 & 0\\
               0  &0\\
\end{array}\right],\,\,\,\,
\hat \Gamma_{R}=\gamma_R \left[ \begin{array}{cc}
              0 & 0\\
              0 &1\\
\end{array}\right],
\label{eq:GaLR}
\eea
and organize the transmission function
\bea
\mathcal T_{L,R}(\epsilon)= {\rm Tr} [\hat \Gamma_L \hat G^r(\epsilon)\hat \Gamma_R \hat G^a(\epsilon) ],
\label{eq:T}
\eea
defined in terms of the retarded and advanced Green’s functions,
$\hat G^r(\epsilon)=[\epsilon \hat I-\hat H+\frac{i}{2}(\hat \Gamma_L+\hat \Gamma_R)]^{-1}$,
$\hat G^a(\epsilon)=[\hat G^r(\epsilon)]^{\dagger}$, $\hat I$ is the identity matrix.
The current leaving the $L$ terminal is calculated from
\bea
I_L=\frac{2e}{h}\int_{-\infty}^{\infty} d\epsilon \mathcal T_{L,R}(\epsilon) [f_L(\epsilon)-f_R(\epsilon)],
\label{eq:ILcoh}
\eea
where $f_{\nu}(\epsilon)=[e^{\beta(\epsilon-\mu_{\nu})}+1]^{-1}$ is
the Fermi function at the $\nu=L,R$ terminal with the chemical potential
$\mu_{\nu}$ and temperature $T=1/(k_B\beta)$, $eV=\mu_{L}-\mu_R$,  $e$ is the electron charge.
In Fig. \ref{Fig2} we display the IV characteristics and the corresponding
rectification ratio
\bea
R(V)\equiv \frac{|I(V>0)|}{|I(V<0)|}
\eea
for the 1,2-bis(4-(phenylethynyl)phenyl)ethane molecule,
reproducing rectification ratios up to $R \sim  1000$ in the coherent limit, as discussed in Ref. \cite{vanderzant15}.

We now assess the role of a thermal environment on this predicted excellent rectifying behavior.
The coupling of electrons to other degrees of freedom, e.g., internal molecular vibrations, solvent, other electrons,
causes phase decoherence of transmitted electrons, inelastic scattering processes, and energy relaxation.
A detailed and accurate description of such processes in molecular conduction
is a challenging task \cite{NitzanRev},
thus for the sake of simplicity and generality, we emulate the environment in a phenomenological manner using
B\"uttiker's probe approach \cite{Buttiker-probe,Pastawski}.
We mimic elastic and inelastic scattering of electrons on the molecule by coupling them to
fictitious reservoirs, termed ``probes".
Specifically, the two molecular sites are coupled to independent (fictitious)
metals, identified by '1' and '2'.
These metals can exchange electrons with the junction, and their properties are determined
self-consistently to introduce relevant thermal effects.
In direct analogy with Eq. (\ref{eq:GaLR}), we construct the molecule-probes hybridization matrices as
\bea &&\hat \Gamma_{1}=\gamma_d\left[ \begin{array}{cc}
              1 & 0\\
               0  &0\\
\end{array}\right],\,\,\,\,
\hat \Gamma_{2}=\gamma_d \left[ \begin{array}{cc}
              0 & 0\\
              0 &1\\
\end{array}\right],
\eea
with $\gamma_d$ the electron-environment coupling strength.
The essential convenience of the probe method lies in its direct correspondence to coherent transport:
The current crossing the junction is calculated using the Landauer-B\"uttiker formula, by
generalizing Eq. (\ref{eq:ILcoh}) to the multi-terminal case,
\bea
I_L=\frac{2e}{h}\sum_{j=R,1,2}\int_{-\infty}^{\infty} d\epsilon
\mathcal T_{L,j}(\epsilon) [f_L(\epsilon)-f_j(\epsilon)],
\label{eq:ILprobe}
\eea
with $\mathcal T_{L,j}(\epsilon)=
 {\rm Tr} [\hat \Gamma_L \hat G^r(\epsilon)\hat \Gamma_j \hat G^a(\epsilon) ]$ 
and the Green's functions generalized to include the molecule-probe coupling,
$\hat G^r(\epsilon)=[\epsilon \hat I-\hat H+\frac{i}{2}(\hat \Gamma_L+\hat \Gamma_R +\hat \Gamma_1 +\hat \Gamma_2)]^{-1}$, $\hat G^a(\epsilon)=[\hat G^r(\epsilon)]^{\dagger}$.
Decoherence and energy exchange processes are implemented with ``voltage probes"  \cite{Buttiker-probe},
by demanding zero net charge flow into each probe,
\bea
I_{1,2}=0,
\label{eq:Vprobe}
\eea
with e.g. the probe-1 current
\bea
I_{1}=\frac{2e}{h}\sum_{j=L,R,2}\int_{-\infty}^{\infty} d\epsilon
\mathcal T_{1,j}(\epsilon) [f_{1}(\epsilon)-f_j(\epsilon)].
\label{eq:I1}
\eea
The functions $f_{1,2}(\epsilon)$ assume a Fermi-function form at temperature $T$
(uniform across the junction), and we search for $\mu_{1,2}$ solving Eq. (\ref{eq:Vprobe}).
Far-from-equilibrium, when the applied bias is large,
an exact analytic solution for this problem is generally
missing, though uniqueness is guaranteed \cite{Jacquet}.
We thus retract to a fully numerical procedure, recently
implemented for electronic conduction \cite{Salil} and phononic heat
transport \cite{Malay, Tulkki}, and employ the Newton-Raphson method by iterating according to
\bea
\begin{bmatrix}
\mu_1^{k+1} \\ \mu_2^{k+1}\\
\end{bmatrix}
=
\begin{bmatrix}
\mu_1^k\\ \mu_2^k\\
\end{bmatrix}
-
\begin{bmatrix}
\ \frac{\partial I_{1}}{\partial \mu_1} && \frac{\partial I_{1}}{\partial \mu_2} \\
\frac{\partial I_{2}}{\partial \mu_1} && \frac{\partial I_{2}}{\partial \mu_2} \\
\end{bmatrix}^{-1}_k
\begin{bmatrix}
\ I_{1} \\ I_{2}\\
\end{bmatrix}_k,
\nonumber\\
\label{eq:iter}
\eea
converging to the unique set of roots for the probes' chemical potentials.
The voltage probe condition (\ref{eq:Vprobe})
allows phase loss processes {\it and} energy exchange. Decoherence without
relaxation can be implemented as well with ``dephasing probes",
by demanding the energy resolved currents at each probe to vanish,
 $I_{1,2}(\epsilon) = 0$. These strict constraints translate into a
set of linear equations for $f_{1,2}(\epsilon)$, which can be readily solved.

The probe approach for decoherence and relaxation has found many applications in mesoscopic
physics for understanding the phenomenology of
charge \cite{Dhar,Udo,Salil} and heat \cite{Malay, Tulkki,DharR,SegalS} transfer.
More recently, several studies had
implemented probes in molecular electronic problems, albeit limited to linear response situations
\cite{Nozaki1,Nozaki2,WaldeckF,Kilgour} or to the simpler dephasing probe approach \cite{Chen-Ratner}.

Below we examine the functionality of tunneling diodes under inelastic effects
by implementing the voltage probe approach. Setting the applied voltage,
we iterate Eq. (\ref{eq:iter}) to convergence,
then calculate the current in the junction from Eq. (\ref{eq:ILprobe}).
Convergence is determined by
ensuring that the probe potentials $\mu_{1,2}$ are evolving monotonically,
$|\mu_{1,2}^{k+1}-\mu_{1,2}^k|\leq|\mu_{1,2}^{k}-\mu_{1,2}^{k-1}|$,
and by confirming that the leakage current to each probe is small,
 $|I_{1,2}/I_L|< 10^{-6}$. Our procedure generally required tens to hundreds of cycles
to converge, with initial conditions at each bias prepared from converged results
at lower voltages.


\begin{figure*}
\vspace{0mm}
{\hspace{-10mm}
\hbox{\epsfxsize=195mm \epsffile{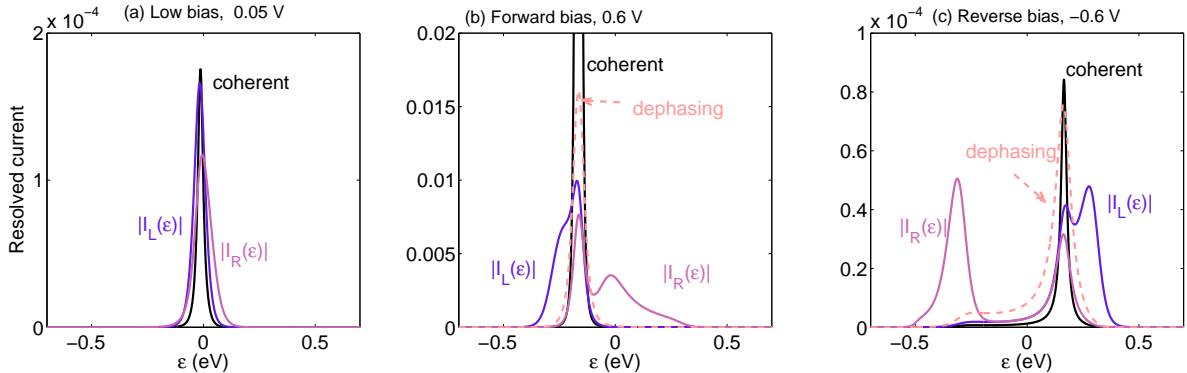}} 
}
\vspace{0mm}
\caption{
Energy resolved currents in
the coherent limit $\gamma_d=0$ (black) and under inelastic effects
$\gamma_d=0.05$ eV. In the latter case $|I_L(\epsilon)|\neq |I_R(\epsilon)|$,
i.e.  the current profile of electrons entering the junction is
different from the one leaving it.
The three panels follow the configurations of Fig. \ref{Fig1}:
(a) Low bias, when inelastic effects play a little role.
(b) At peak voltage  $\sim 0.6$ V the two states are in resonance,
$\epsilon_1+\alpha eV/2\sim \epsilon_2-\alpha eV/2=-0.165$ eV.
(c) At large negative voltages $-0.6$ V,  $\epsilon_1+\alpha eV/2$=0.165 eV and $\epsilon_2-\alpha eV/2=-0.49$ eV.
In panels (b)-(c) we further depict
the resolved current under the dephasing probe condition
when energy relaxation is not allowed (dashed line).
Parameters are the same as in Fig. \ref{Fig2}.
}
\label{Fig3}
\end{figure*}

\section{Results}

\subsection {Two-state diodes}

The functionality of the junction as a diode is demonstrated in Fig. \ref{Fig2}.
Inelastic-environmental effects deteriorate the rectifying operation, as expected, yet
the junction rectifies substantially even when $\gamma_d$ is order of the tunneling energy $v$.
When inelastic effects are strong, $\gamma_d\gg v$, the junction
supports only $R \sim 50$. 
Naturally, these values reduce when $\epsilon_2$ is placed closer to
the Fermi energy, when the gap $\epsilon_1-\epsilon_2$ is
reduced, and when $\gamma_{L,R}$ takes larger values, as was demonstrated in Ref. \cite{vanderzant15}.

To understand the decline in the diode operation under environmental effects we display in Fig. \ref{Fig3}
the energy resolved currents at the two terminals,
$I_L(\epsilon)$ and $I_R(\epsilon)$.
These currents are calculated from the integrand in Eq. (\ref{eq:ILprobe})
and the corresponding equation for the right terminal. We find that at small biases
($0.05$ V) the resolved current arranges a single
peak located close to the Fermi energy. The role of the
environment is to broaden this peak,
and to slightly enhance the current due to the contribution of inelastic terms.
The behavior of the resolved currents at the voltage $0.6$ V is displayed in Fig. \ref{Fig3}(b). This
voltage arranges a resonance configuration most beneficial for conductance, see Fig. \ref{Fig1}(b).
When $\gamma_d=0$, a single-large contribution
to the current shows up (clipped here for clarity), located around $\epsilon$ = -0.165 eV,
corresponding to the energy of the two molecular sites.
When we allow energy exchange with the environment,
electrons with energies reaching $\mu_R= 0.3$ eV contribute, see
$|I_R(\epsilon)|$. These electrons dissipate their energy on the junction, and are re-emitted
at the other terminal with lower energies,
down to $\mu_L=-0.3$ eV, see the behavior of $|I_L(\epsilon)|$.
Under the reversed bias  $-0.6$ V,
corresponding to the configuration of Fig. \ref{Fig1}(c),
an analogous behavior takes place, see Fig. \ref{Fig3}(c).


We further present in Figs. \ref{Fig3}(b)-(c)
the resolved current under the dephasing probe condition, when decoherence effects take place
without relaxation.
In this case, energy conservation is enforced and electrons on the junction do not absorb/dissipate
energy from/to the thermal bath, thus
$|I_L (\epsilon)| = |I_R (\epsilon)|$. The effect of the environment 
is to broaden the coherent current profile, again disturbing the
operation of the diode. Interestingly, the microscopic differences between the two probes as reflected in Fig.
\ref{Fig3} do not translate to the macroscopic operation of the device in the present model,
and the total-integrated currents under either probes are almost identical.
This observation however is not general, and
deviations between probes (emulating different microscopic processes) may clearly affect functionality,
particularly in multi-site chains.

\begin{figure}[htbp]
\vspace{0mm}
\hspace{-0mm}{\hbox{\epsfxsize=85mm \epsffile{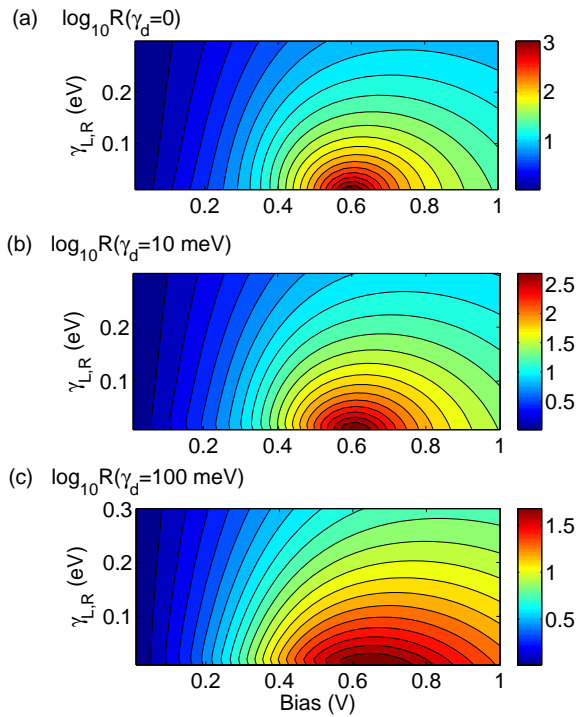}}} 
\vspace{-1mm}
\caption{
Map of rectification ratios as a function of applied bias and metal-molecule hybridization.
(a) Coherent limit, (b) $\gamma_d=10$ meV, and (c) $\gamma_d=100$ meV.
Parameters are the same as in Fig. \ref{Fig2}.
}
\label{Fig4}
\end{figure}

The departure from the strict tunneling picture limits the operation of the junction as a diode
since the contrast between resonant transmission and tunneling no longer
stringently holds with inelastic electrons contributing to the current.
Note that $\gamma_{L,R}$, the coupling of the molecule to the metals, corresponds to levels' broadening, and
it affects the diode in a parallel manner as the tunneling current becomes
over-dominated by ballistic electrons.
This effect is presented in Fig. \ref{Fig4}, displaying contour plots (log scale)
of rectification ratios as a function of $\gamma_{L,R}$ and applied
voltage bias at different values for $\gamma_d$.

\begin{figure}[htbp]
\vspace{0mm}
\hspace{-0mm}{\hbox{\epsfxsize=80mm \epsffile{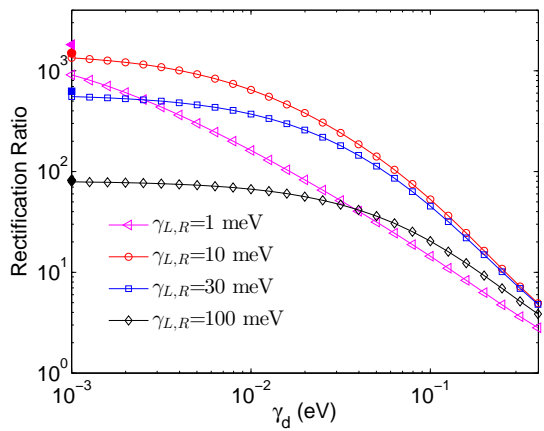}}} 
\vspace{-1mm}
\caption{
Decline of rectifying behavior due to environmental
interactions with the parameters of Fig. \ref{Fig2}.
Filled symbols at the left boundary mark
corresponding rectification ratios in the coherent $\gamma_d=0$ limit.
}
\label{Fig5}
\end{figure}

In Fig. \ref{Fig5} we further present the rectification ratio at peak voltage, $R(V_p)$, as
a function of $\gamma_d$. 
Weakly hybridized junctions operate better
as tunneling diodes, yet they are more susceptible to inelastic effects.
When the levels are largely broadened with $\gamma_{L,R}$ = 0.1 eV,
the junction still rectifies well
with $R \sim 100$ in the coherent limit,
and it is resistant to the environment
as long as $\gamma_d \lesssim 0.1$ eV.

\subsection{Four-state diodes: sites with quasidegenerate states}

\begin{figure}[htbp]
\vspace{0mm}
\hspace{-0mm}{\hbox{\epsfxsize=90mm \epsffile{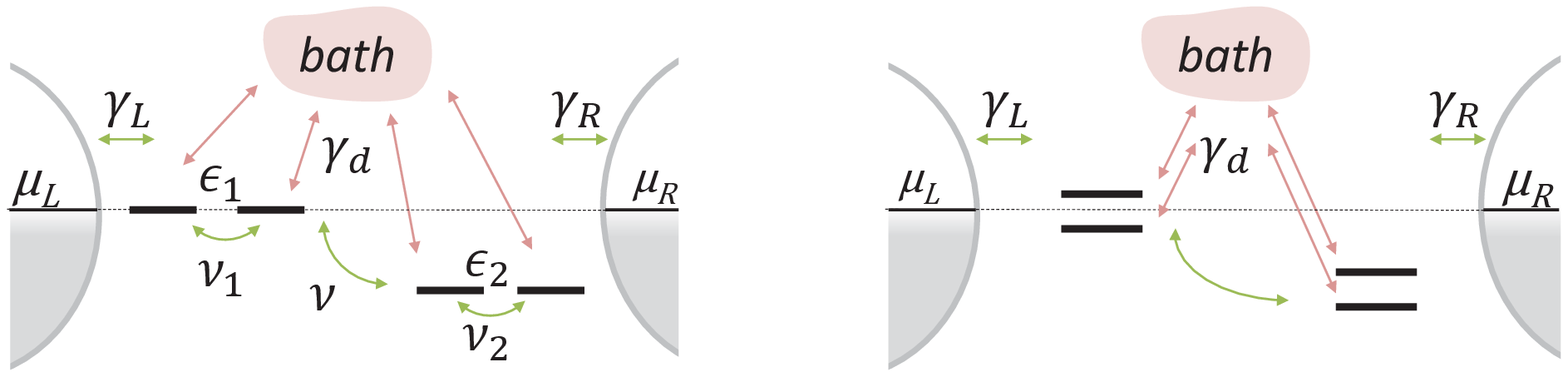}}} 
\hspace{-13mm}{\hbox{\epsfxsize=90mm \epsffile{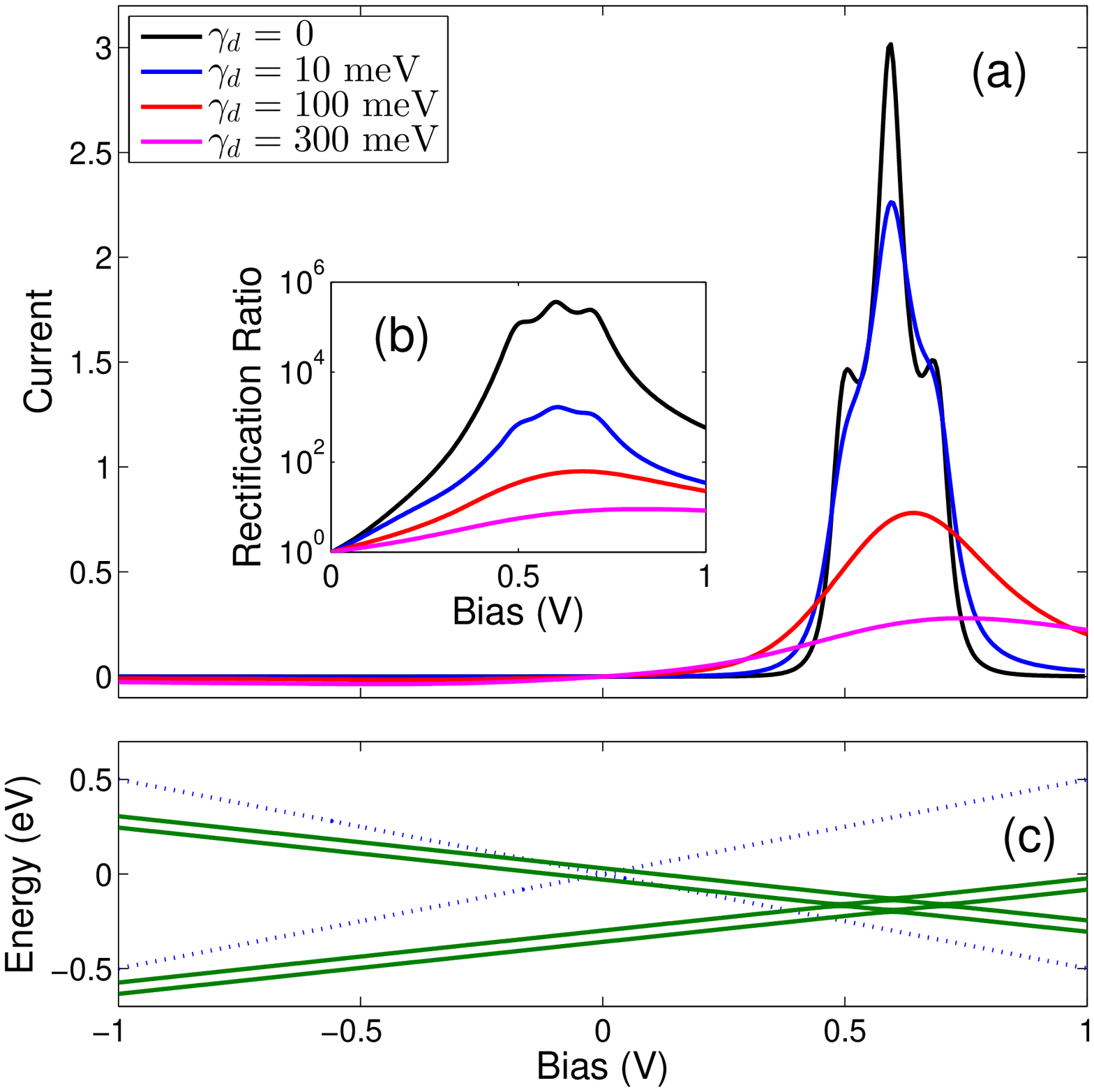}}} 
\vspace{-1mm}
\caption{
Top: Scheme of a four-site diode 
in the site representation (left), and energy representation within each conjugated moiety (right).
Bottom: Operation of the junction as a diode
(a) IV characteristics, (b) rectification
ratio, and (c) level diagram with (dashed lines) chemical
potentials and (full lines) the four eigenenergies of the
molecule.
Parameters are $\epsilon_1=0$, $\epsilon_2=-0.329$, $v=0.0109$, $\gamma_{L,R}=0.03$ eV,
and $\alpha=0.55$ as in Fig. \ref{Fig2},
as well as $v_{1}=v_2=0.03$ eV for the tunneling energy within each conjugated unit.
The temperature is set at $T=298$ K and
$\gamma_d$ was varied as indicated in the figure.
}
\label{Fig6}
\end{figure}

The rectifying behavior of a tunneling diode 
could be enhanced significantly if we mange to attenuate
the tunneling current under reverse bias,
while keeping intact the resonant current in the forward direction.
We recall that in an $N$-site molecular chain the tunneling conductance decays exponentially
with distance, while the resonant-coherent
current is independent of chain length \cite{NitzanBook}
(leaving aside decoherence and dissipation effects).
We thus suggest to generalize the two-state tunneling diode of Fig. \ref{Fig1}
and include four molecular states, two sites within each conjugated unit, see Fig. \ref{Fig6}.
In the energy representation,
each conjugated unit includes two closely spaced orbitals
(at zero bias) instead of a single level.
This principle was employed in Ref. \cite{Nijhuis15} to experimentally realize highly rectifying junctions,
albeit employing only one conjugated unit and a long saturated tail.
The Hamiltonian of the four-site molecule takes the form
\bea \hat H=\left[  \begin{array}{cc}
\hat H_1
& \hat V_{12}\\
\hat V_{21} & \hat H_2 \\
 \end{array}\right],
\nonumber
\label{eq:H4}
\eea
with $\hat H_{1,2}$ describing the conjugated units and $\hat V_{12}$ and  $\hat V_{21}$ comprising the
tunneling terms through the $\sigma$ bridge,
\[ \hat H_1= \begin{pmatrix}
\epsilon_1+ \alpha eV/2 & v_1 \\ v_1 & \epsilon_1+ \alpha eV/2 \end{pmatrix}; \,\,\,
\ \hat V_{12}=\begin{pmatrix}  0 &0   \\ v & 0 \end{pmatrix};  \,\,\,\
\]
\[
\ \hat H_2=\begin{pmatrix}  \epsilon_2- \alpha eV/2 & v_2   \\ v_2& \epsilon_2- \alpha eV/2 \end{pmatrix};  \,\,\,\
\hat V_{21}=\begin{pmatrix}  0 &v   \\ 0 & 0 \end{pmatrix}.  \,\,\,\
\]
The two orbitals within each conjugated unit are made
close in energy; $v_{1,2}$ determines departure from degeneracy.

To make a meaningful comparison to the two-site junction, we employ here the same parameters as in Fig. \ref{Fig2}.
The IV characteristic of the model is displayed in
Fig. \ref{Fig6}, demonstrating a very large rectification ratio in the coherent limit,
but a strong decline in operation under environmental effects.
The four-site design thus offers an improved diode functionality, yet a greater sensitivity to
thermal effects since a minuscule tunneling current is more easily outplayed by inelastic processes.

\section{Conclusions}

We analyzed the operation of molecular tunneling
diodes comprising conjugated-saturated-conjugated segments under environmental effects inducing
decoherence and energy relaxation. Specifically, we employed a two-site molecular junction with parameters
corresponding to the 1,2-bis(4-(phenylethynyl)phenyl)ethane junction \cite{vanderzant15}.
This system operates as a diode since (i) an asymmetry is built into the molecular backbone,
 and (ii) an internal potential drop develops within the molecule in response to applied bias. 
For these reasons, we expect our findings are fully general to constructions with different backbones
or system energetics, provided they follow these design principles.
The contrast between resonant and tunneling conductances is the key to the
optimization of the diode, with rectification ratios reaching three orders of magnitude in coherent situations.
The diode performs well under environmental effects as long as
$\gamma_d \lesssim v$, providing $R\sim 100-500$.
The decline in the operation of the diode under thermal effects ensues
from the contribution of
incoherent electrons, playing down the contrast between resonant transmission and tunneling conductance. We
further proposed the design of an improved rectifier, suggesting a chemical backbone supporting two closely spaced orbitals within
each conjugated unit. This structure provides $R\sim 10^5$
within our parameters when $\gamma_d=0$,
though this type of diode shows a greater susceptibility to environmental effects.

%

An inelastic vibration-assisted diode obeying a similar principle
of level alignment was recently investigated in Refs. \cite{SiminePCCP, SimineJCP},
with a weak, phonon-assisted hopping energy instead of the direct tunneling $v$.
Despite the different conduction mechanism (resonant vs. hopping), this setup produced
IV characteristics closely resembling the curve in Fig. \ref{Fig2}.

The probe technique employed here for mimicking decoherence and inelastic effects
can be feasibly implemented within DFT + NEGF calculations to receive first
indications on the role of thermal effects on electron transport in nanojunctions \cite{Nozaki1,Nozaki2}.
Our work here is significant beyond the specific analysis of tunneling diodes as we demonstrate for the first time the
implementation and utility of the voltage probe method to molecular electronic problems
under {\it large} biases, introducing
inelastic effects into far-from-equilibrium molecular conduction.
Our approach allows the interrogation of the role of the environment on
other elemental molecular electronic functionalities including
negative differential conductance \cite{dubi}, single-molecule transistors \cite{transistors},
and nonlinear, phonon-assisted, thermoelectric energy conversion \cite{JianHua}.

\begin{acknowledgments}
This work was funded by a Discovery Grant from the
Natural Sciences and Engineering Research Council of
Canada and the Canada Research Chair Program.

\end{acknowledgments}




\begin{thebibliography}{99}
\bibitem{commentMB}
Asymmetry and many body interactions are necessary but
insufficient conditions for demonstrating the diode effect. In some
models only specific types of spatial asymmetry resulted in the
diode effect, see e.g. \cite{Roy,Collet}.
\end{thebibliography}
\end{document}